\documentclass[pdflatex,sn-basic,iicol]{sn-jnl}
\setcitestyle{numbers, square,open={[},close={]}}

\jyear{2023}%

\theoremstyle{thmstyleone}%
\theoremstyle{thmstyletwo}%

\theoremstyle{thmstylethree}%

\begin{document}

\title[Measuring the diversity of metadata]{Measuring the diversity of data and metadata in digital libraries}


\author*[1]{\fnm{Rafael} \sur{C. Carrasco}}\email{carrasco@ua.es}

\author[1]{\fnm{Gustavo} \sur{Candela}}\email{gcandela@ua.es}

\author[1]{\fnm{Manuel} \sur{Marco-Such}}\email{marco@dlsi.ua.es}

\affil*[1]{\orgdiv{Departmento de Lenguajes y Sistemas Informáticos}, \orgname{Universidad de Alicante}, \orgaddress{\street{Carretera San Vicente del Raspeig s/n}, \city{San Vicent del Raspeig}, \postcode{03690}, \state{Alicante}, \country{Spain}}}

\abstract{Diversity indices have been traditionally used to capture the biodiversity of ecosystems by measuring the effective number of species or groups of species. In contrast to abundance, which is correlated with the amount of data available, diversity indices provide a more robust indicator on the variability of individuals. These types of indices can be employed in the context of digital libraries to identify trends in the distribution of topics, compare the lexica employed by different authors or analyze the coverage of semantic metadata.}




\maketitle

\section{Introduction}\label{sec1}

\emph{Richness}, usually defined as the number of species present in an ecosystem, provides a limited picture of its biodiversity as it weights all groups equally, regardless their relative abundances. In contrast, \emph{diversity indices}~\cite{Hill1973} are numerical estimators that measure both richness and evenness by giving more
relevance to abundant species. They therefore provide an effective number of species which is more robust than the sample size, due to the smaller contribution of rare, possibly undetected, cases.

As digital libraries become more readily available, there is an increasing need to explore which bibliometric measures could make their features easier to understand. It has been argued \cite{Kyle2021} that diversity indices
could effectively disentangle the correlation between richness and data volume. The purpose of this paper is therefore to analyze how diversity indices could assist researchers and professionals in evaluating the lexical diversity of the content as well as the metadata
coverage in digital collections.

As regards textual content, the \emph{type-token ratio} (TTR) has been traditionally employed to measure the lexical diversity of documents. The TTR is computed as the number of different words (types) divided by the number of words (tokens) in the text. For example, previous works compare different approaches, including MLTD \cite{McCarthy2010} and \emph{vocd}~\cite{McKee2000}, to evaluate TTR and its variability within a sample. Some researchers~\cite{Kubat2013} have also explored whether genres could be characterized by specific TTR probability distributions.

Previous research has suggested applying diversity indices to evaluate the lexical richness of documents ~\cite{Jarvis2013}. But other features of digital libraries could also benefit from analysis using diversity
concepts. For example, the local and temporal variations in the coverage of topics or authors could be better examined by computing diversity indices, as they are not as sensitive to infrequent items which are not representative of the collection.

Let us recall that, in ecology, the true diversity, or diversity index of order $k$ for an ecosystem with $N$ groups or species, is defined as
\begin{equation}
  \label{eq:diversity}
  D^{[k]} = \left(\sum_{n=1}^N  p_n^k\right)^{\frac{1}{1-k}}
\end{equation}
where $p_n$ is the probability or relative abundance of the $n$-th
class, and the parameter $k$ determines the relative weight of frequent
versus infrequent groups: the larger $k$ is, the less significant rare
species are.

There is therefore a family of indices $D^{[k]}$, the Shannon
index ($k=1$) and the Simpson index ($k=2$) among the most
popular~\cite{Roswell2021}. Although the parameter $k$ influences
the value of the diversity obtained, the exact choice is not critical
when the objective is to compare diversities at different locations or
time intervals. In particular, when addressing digital library data
and metadata, $k=1$ becomes a natural choice, as $D^{[1]}$ can be
easily connected to the \emph{entropy} of a
source~\cite{Shannon1948},  defined in information theory
as
$$
H = -\sum_{n=1}^N p_n \log p_n
$$
It is thus not difficult to prove that, as $k$ approaches 1, one obtains $D^{[1]} = \exp(H)$. We also note that $k=0$ leads to the richness $R$ of the sample.

In this paper we will explore the applicability of diversity indices to analyzing data (Section~\ref{sec:lexical}) and metadata (Section~\ref{sec:metadata}) produced by digital libraries. Our comparison between libraries will be based on \emph{linked open data} collections~\cite{BernersLee2006} published by libraries, as they provide an open benchmark.

\section{Lexical diversity}
\label{sec:lexical}
\begin{figure*}[h]
  \centering
  \includegraphics[width=0.8\linewidth]{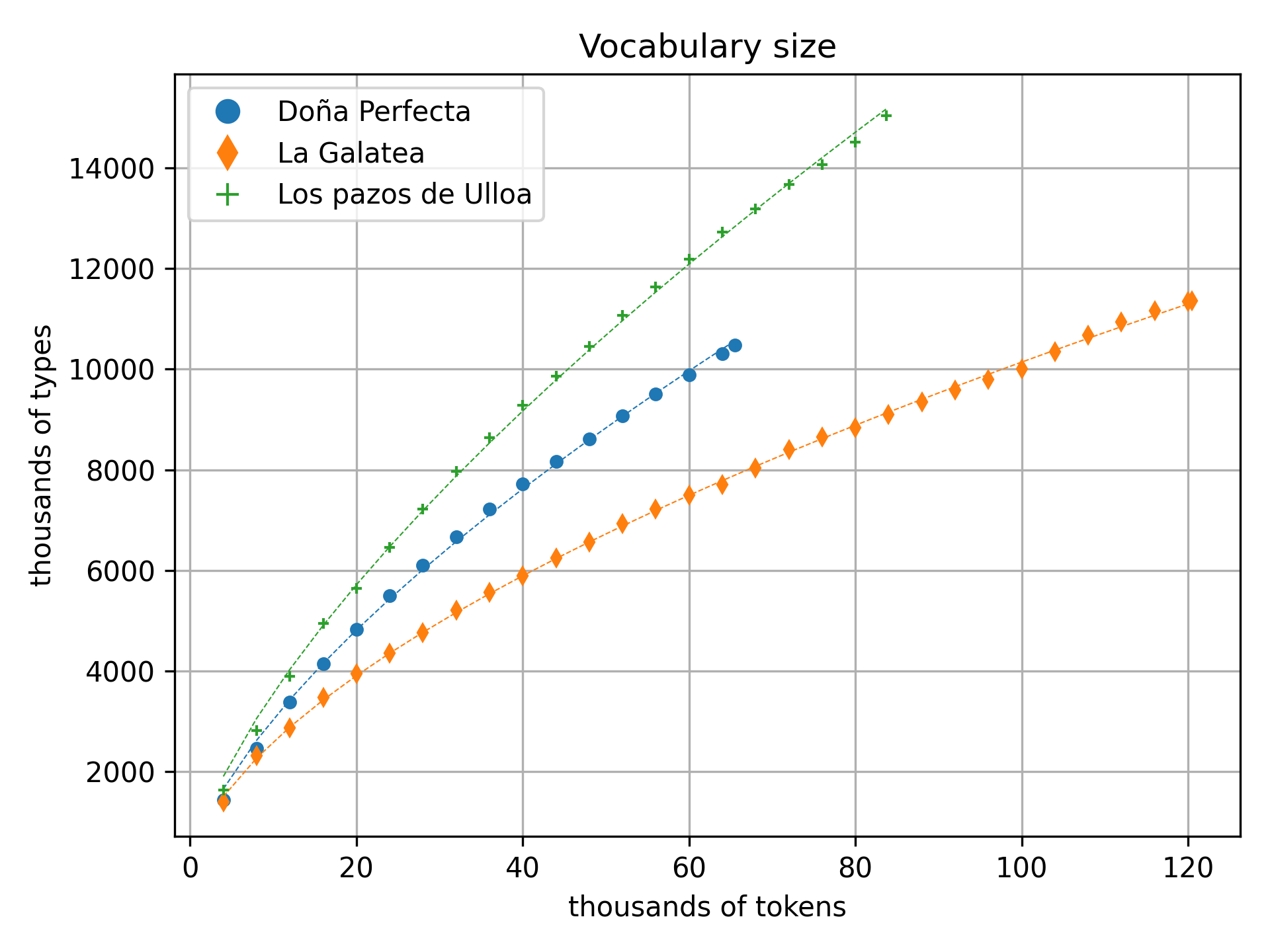}
  \caption[Dictionary size]{Vocabulary size as a function of the number of tokens read for three novels: \emph{Los pazos de Ulloa} by Emilia Pardo Bazán, \emph{Doña Perfecta} by Benito Pérez Galdós and \emph{La Galatea} by Miguel de Cervantes Saavedra.}
  \label{fig:size_of_dict}
\end{figure*}

The number $M$ of entries in its vocabulary, also known as the number of token types, provides an indication of the lexical diversity of a document. The number of token types depends, however, on the document length, and $M$ shows a monotonous growth with the number $n\le N$ of
tokens processed, $N$ being the document length (see
Figure~\ref{fig:size_of_dict}). This unbounded growth is consistent with the well known fact that tokens in a collection approximately follow a Zipfean distribution~\cite{Piantadosi2014}. However, this impedes
a direct comparison of texts based on the size of the vocabulary used.

The number of token types in the plots can be accurately approximated by a power function $C n ^\alpha$ with only two parameters: the scale $C$ and the exponent $\alpha$. The parameters that best fit the examples can be found in Table~\ref{tab:size_of_dict_params}, and they have been used to draw the lines in Figure~\ref{fig:size_of_dict}, which closely follow the data points.

\begin{table}[h]
  \centering
  \begin{tabular}{lrr}
    & $C$ & $\alpha$ \\\hline
    Los pazos de Ulloa & 6.7 & 0.68 \\
    Doña Perfecta & 6.9 & 0.66\\
    La Galatea & 11.1 & 0.59\\\hline
    \\
  \end{tabular}
  \caption{Optimal parameters for the lines $Cn^\alpha$ depicted in
    Figure~\ref{fig:size_of_dict}.}
\label{tab:size_of_dict_params}
\end{table}

\begin{figure*}
  \centering
  \includegraphics[width=0.8\linewidth]{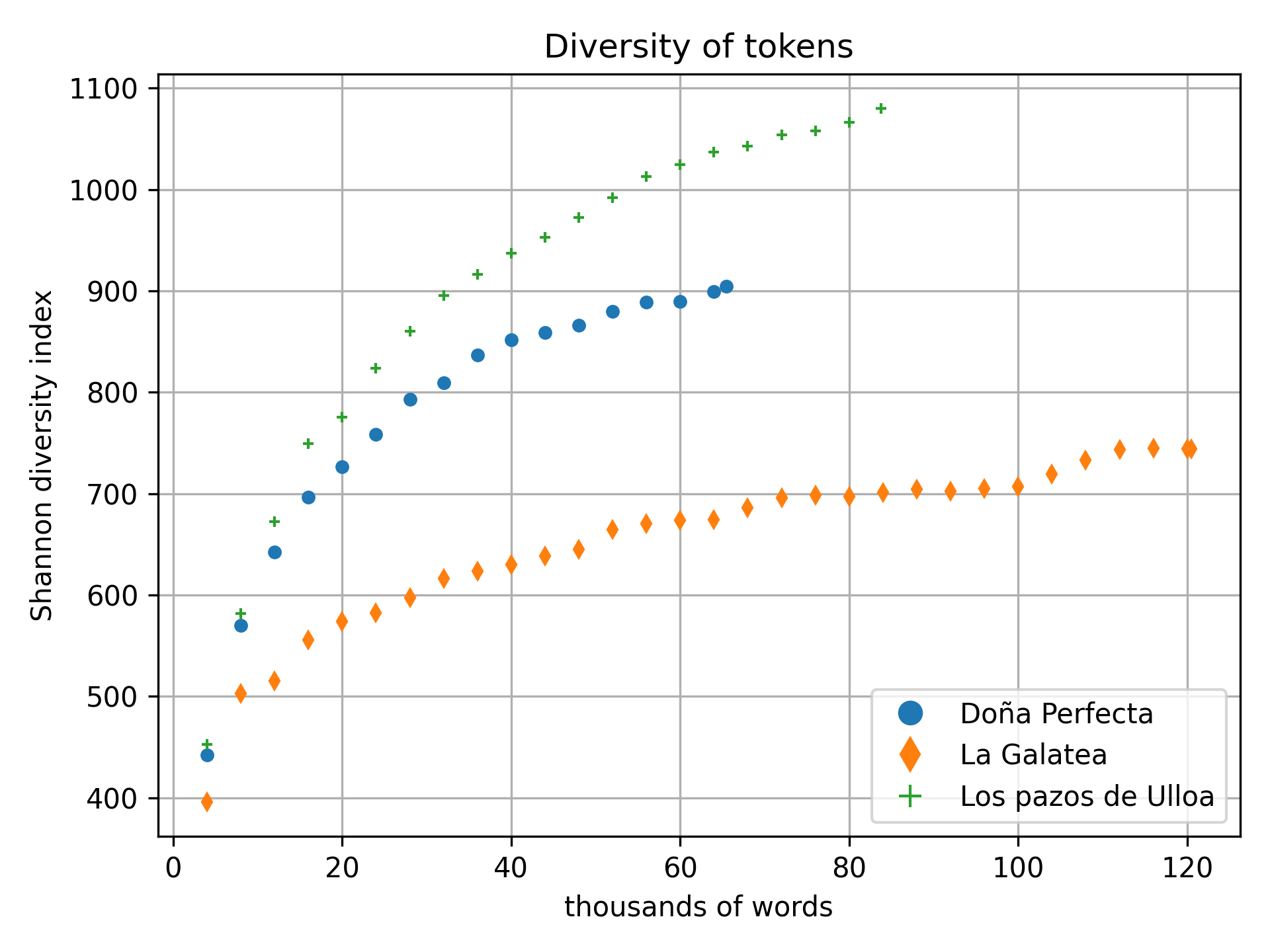}
  \caption[models]{Shannon diversity index for the works presented in
    figure~\ref{fig:size_of_dict}.}
  \label{fig:models}
\end{figure*}

A potential advantage of diversity indices is that they consist of a single finite value with an intuitive interpretation. The diversity of types can be calculated exactly if the underlying probability distribution of the vocabulary is known (and stationary), but, in practice, the probabilities must be estimated from a text sample
using the observed frequencies instead. As the accuracy of the estimation increases with the text length, the result will converge to the true value as the number of tokens grows. In the most common situation, however, the sample size is not large enough to approximate the asymptotic value: as shown in Figure~\ref{fig:models}, the Shannon
diversity index is usually still growing when the end of the document is reached.

The diversity plots in Figure~\ref{fig:models} call for a saturating function to model the observed shape. A function which has been traditionally used to estimate biodiversity from samples of variable
size~\cite{Colwell1994} is the saturating exponential
\begin{equation}
  \Delta_{M1}(n) = D\; (1 - e ^{\alpha n}),
\end{equation}
which involves only two parameters, the exponent $\alpha$ and the asymptotic value $D$ of the diversity index.

A second traditional asymptotic model~\cite{Colwell1994} for species accumulation curves is the two-parameter function
\begin{equation}
  \Delta_{M2}(n) = D\; \frac{n}{n + c}.
\end{equation}

In our experiments, when models M1 and M2 were extrapolated, they usually underestimated the diversity of larger samples. We therefore investigated additional saturating functions, in particular, a generalized quotient of monomials
\begin{equation}
  \Delta_{M3}(n) = D \left(\frac{n + b}{n + c}\right),
\end{equation}
and the powered quotient
\begin{equation}
  \Delta_{M4}(n) = D \left(\frac{n}{n + c}\right)^\alpha.
\end{equation}
We note that in all models, $D$ is the asymptotic value, that is, the true diversity index.

\begin{figure*}
  \centering
  \includegraphics[width=0.8\textwidth]{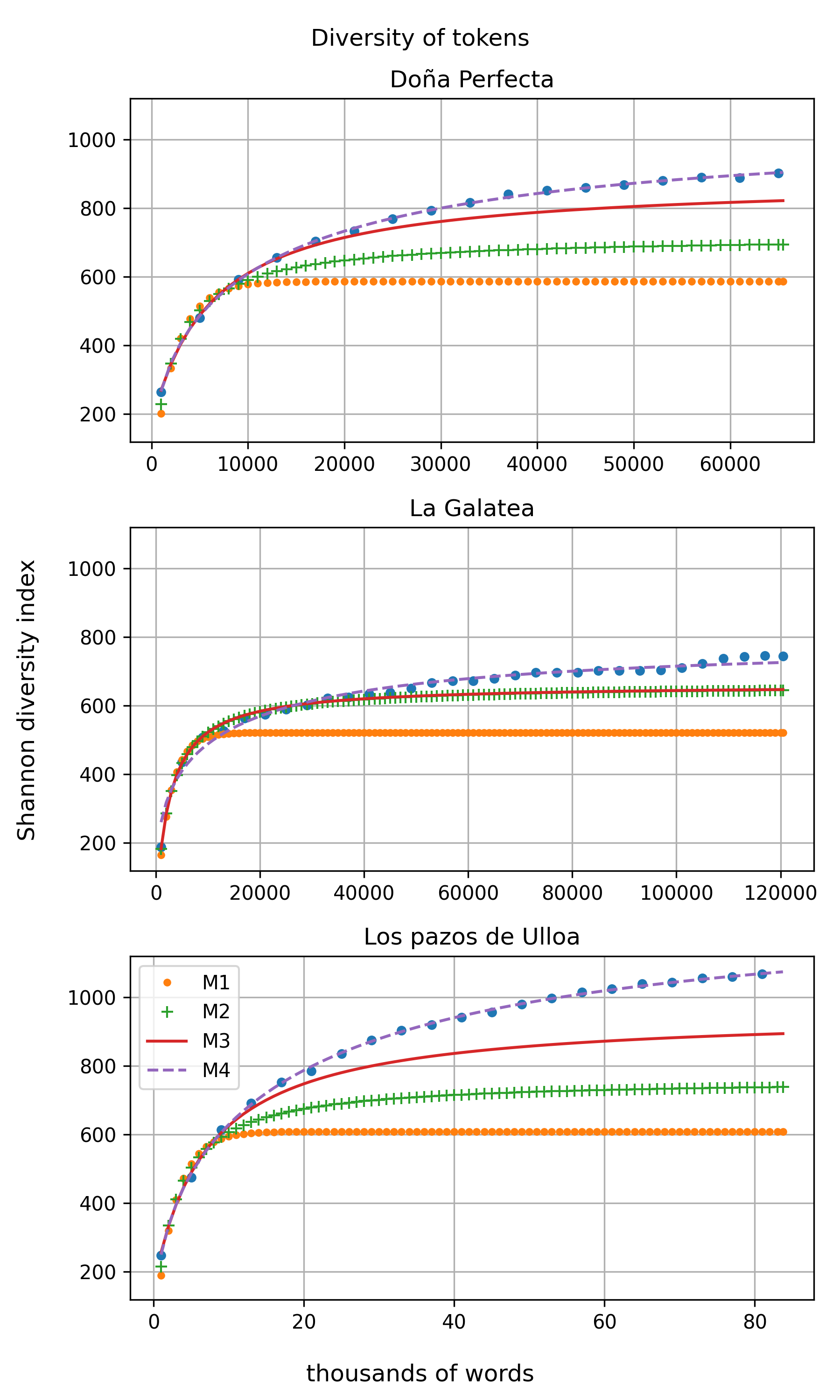}
  \caption{Predictive power of the models when the initial 10000 tokens are used to identify the optimal parameters.}
  \label{fig:predict}
\end{figure*}

When ten thousand tokens were used to extrapolate the curve for larger values, the results showed that model M4 consistently outperformed the others (see Figure~\ref{fig:predict}). It can be argued that, given
the high accuracy of the predictions, the extrapolated diversity computed by model M4 (the value of parameter $D$) can be used to compare the lexical diversity of texts or that of collections labeled by author, genre or historical period.

\begin{figure*}
  \centering
  \includegraphics[width=0.9\textwidth]{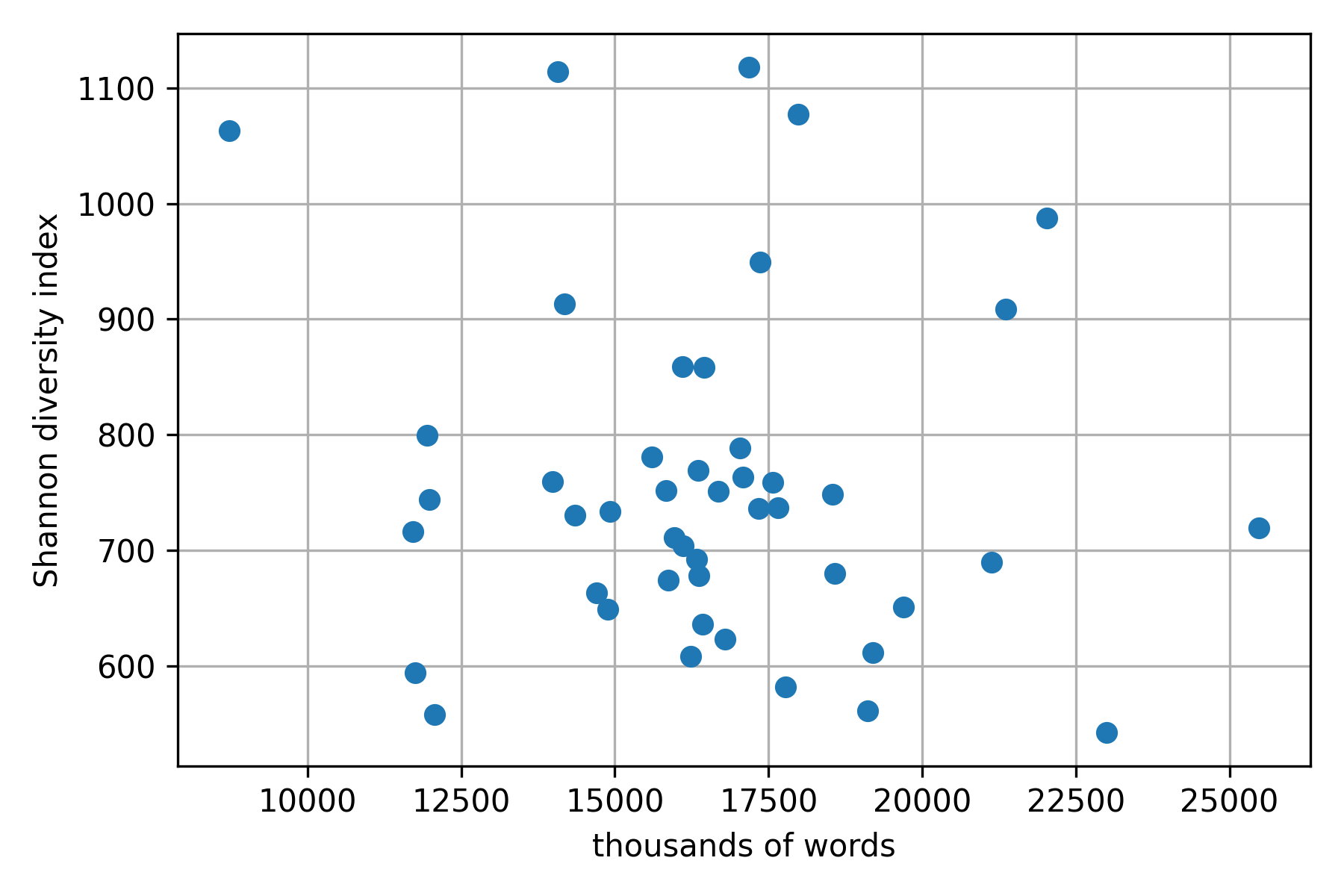}
  \caption{Shannon diversity index of books by Lope de Vega.}
  \label{fig:Lope}
\end{figure*}

Our results show that the value predicted with model $M4$ does not depend on the size of the sample text. As an illustration, Figure~\ref{fig:Lope} shows the lexical diversity of works by a prolific author (Lope de Vega) as a function of the text length. The variability we found could be associated with the style of the work (for
example, works with rhyming tend to exhibit higher diversity), but the diversity has no significant correlation with the length of the work (Pearson's $R\simeq -0.08$).

\section{Metadata diversity}
\label{sec:metadata}
\subsection{Catalographic records}
\label{sec:marc}
\begin{figure*}
  \centering
   \includegraphics[width=0.45\textwidth]{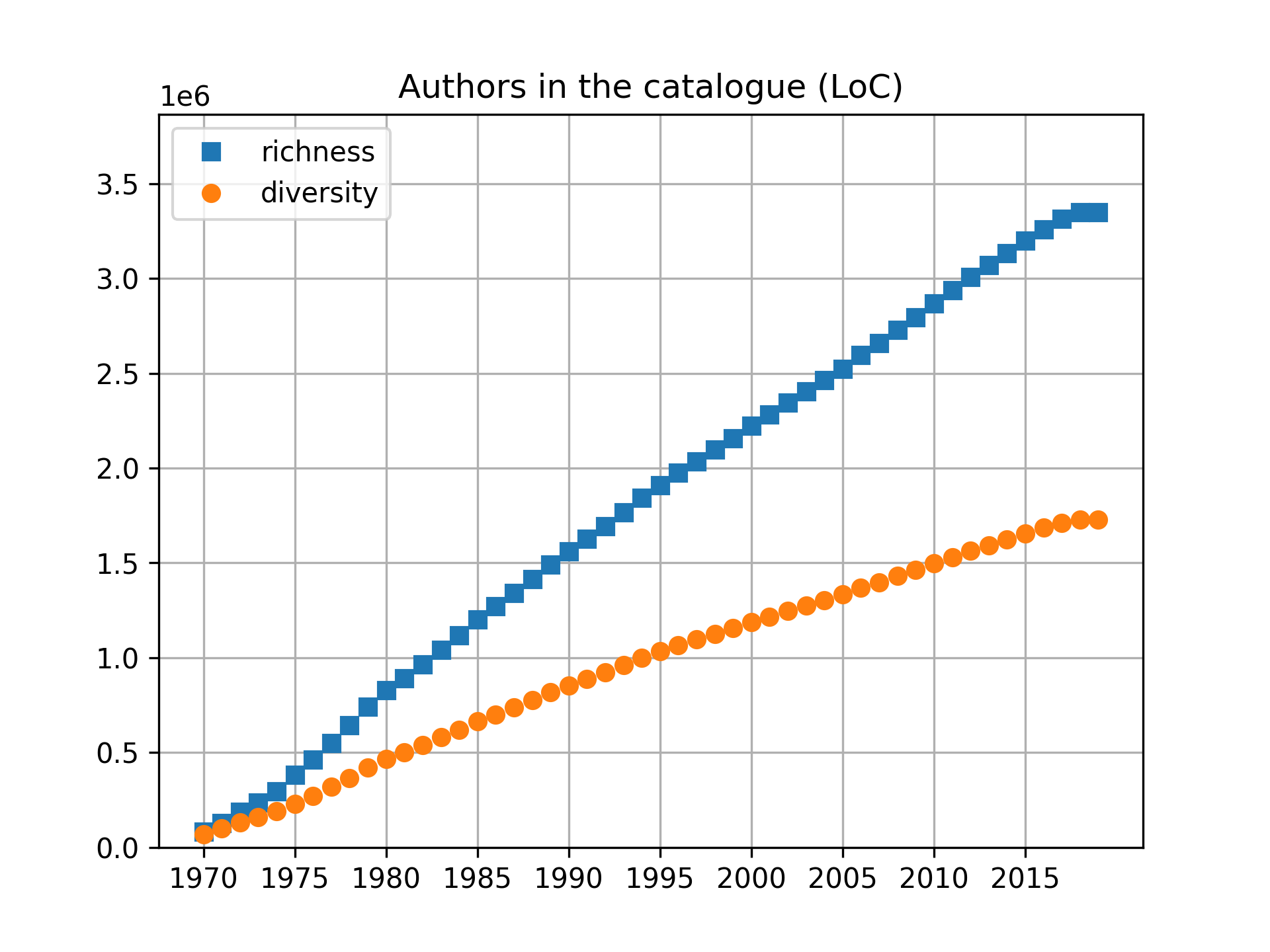}
   \includegraphics[width=0.45\textwidth]{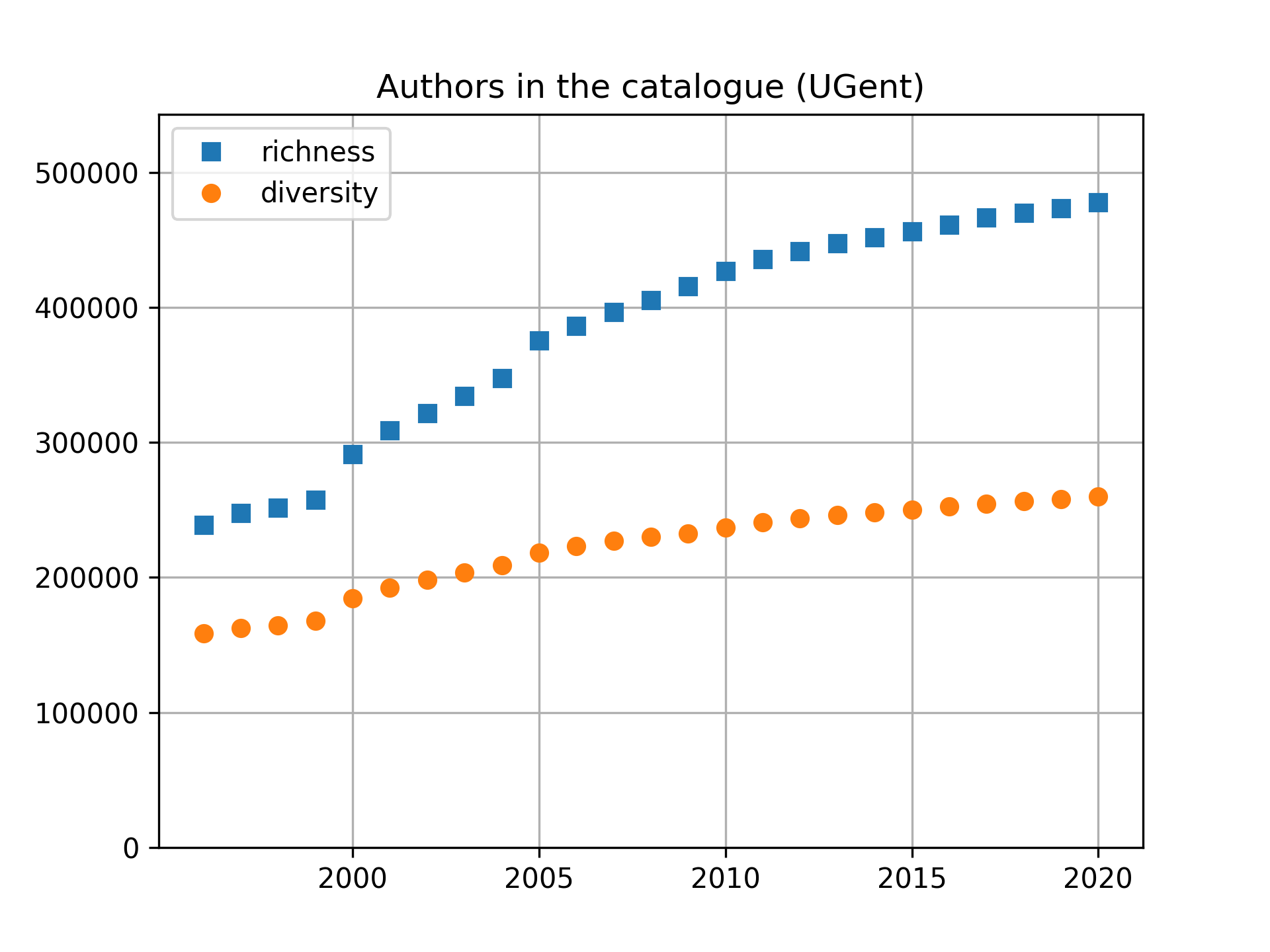}
   \includegraphics[width=0.45\textwidth]{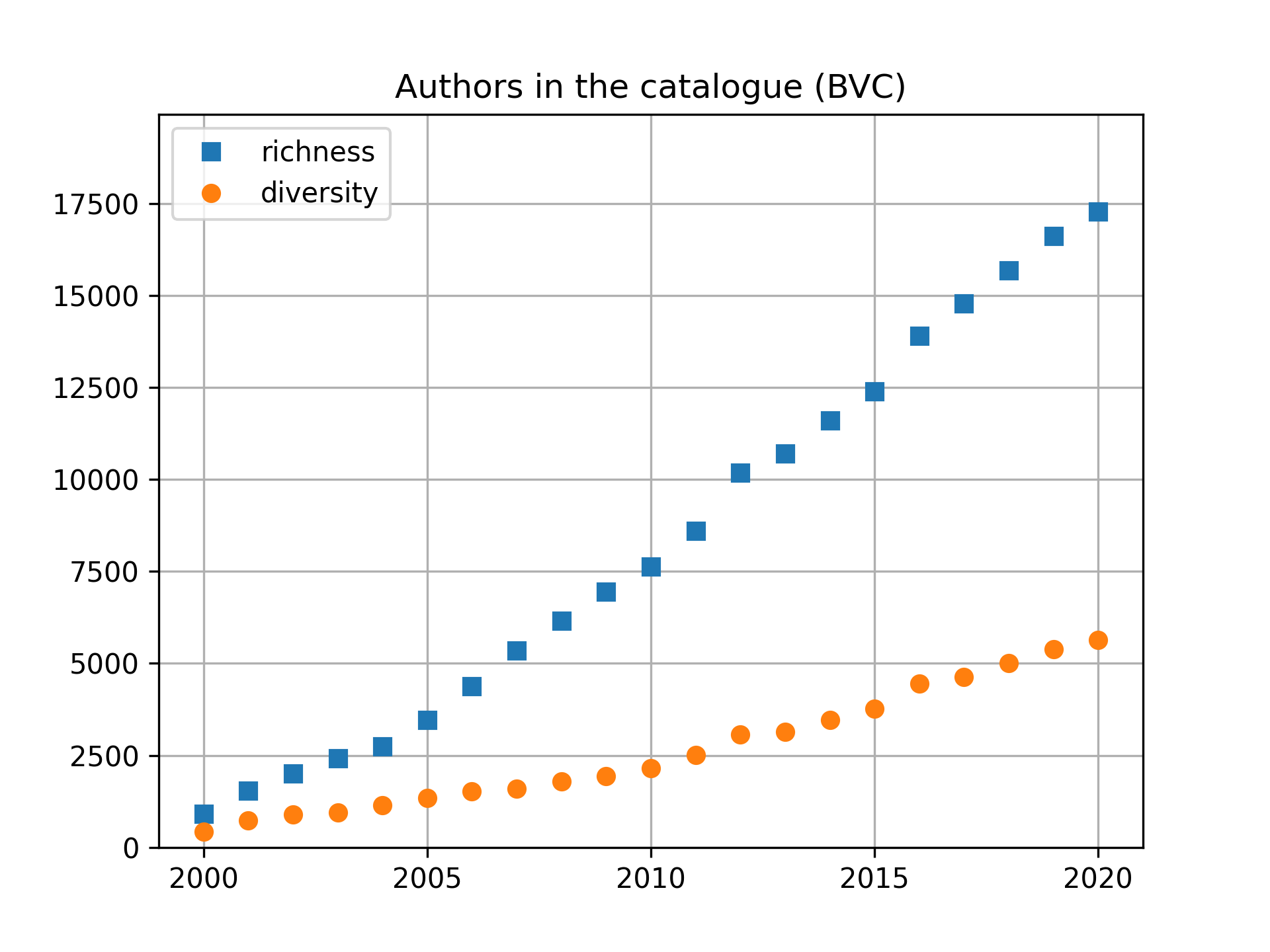}
   \caption{Cumulative number of authors and Shannon diversity of the authors in the catalog as a function of the year the MARC record entered the catalog.}
  \label{fig:authors}
\end{figure*}

Diversity indices can be also employed to analyze the catalographic metadata created by digital libraries. For example, Figure~\ref{fig:authors} shows the richness and diversity of book authors in the catalogs of three libraries which have published comprehensive collections of catalographic data using open licenses: a large library (Library of Congress, LoC\footnote{Library of Congress
 full book records: \url{www.loc.gov/item/2020445551}}), a medium-sized library (Universiteitsbibliotheek Gent, UGent\footnote{University of Gent book records: \url{lib.ugent.be/info/exports}}), and a small library
(Biblioteca Virtual Miguel de Cervantes, BVC\footnote{Miguel de Cervantes book records:
  \url{data.cervantesvirtual.com/datasets}}).

The richness and diversity lines show a monotonic growth over time with no indication that a plateau could be reached soon. The smaller ratio between diversity and richness for the BVC library (about 33\%) in comparison to the ratio for the LoC and UGent collections (52--54\%) is a reflection of its narrower scope---the BVC focuses on
Hispanic literature and history---which shows a reduced fraction of the authors providing a vast contribution to the catalog. Indeed, the average number of items per author in the BVC collection is $\mu=4.9$, while this average is lower for the LoC ($\mu=2.5$) and UGent
library ($\mu=2.1$).

\begin{figure*}
  \centering
  \includegraphics[width=0.45\textwidth]{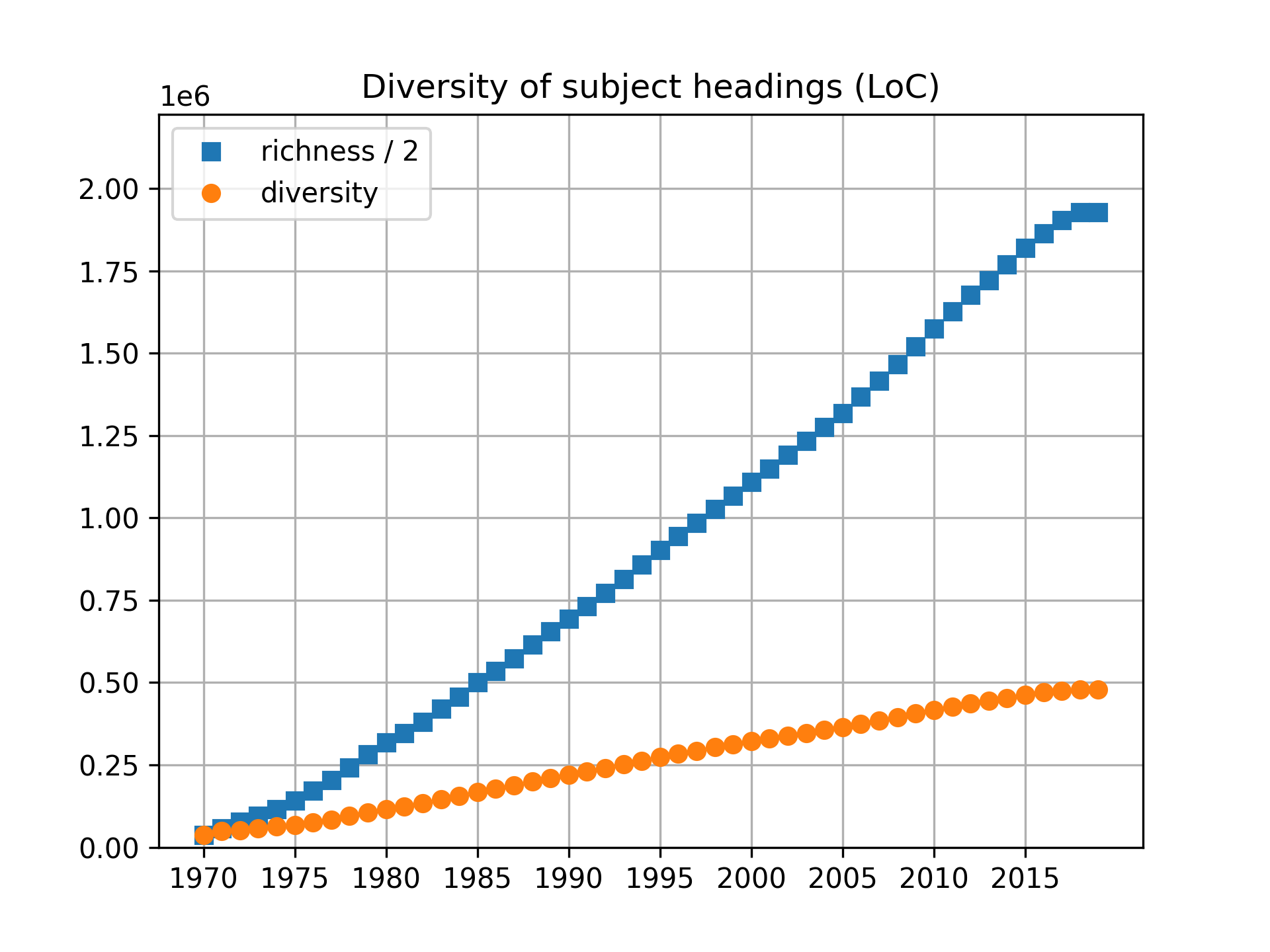}
  \includegraphics[width=0.45\textwidth]{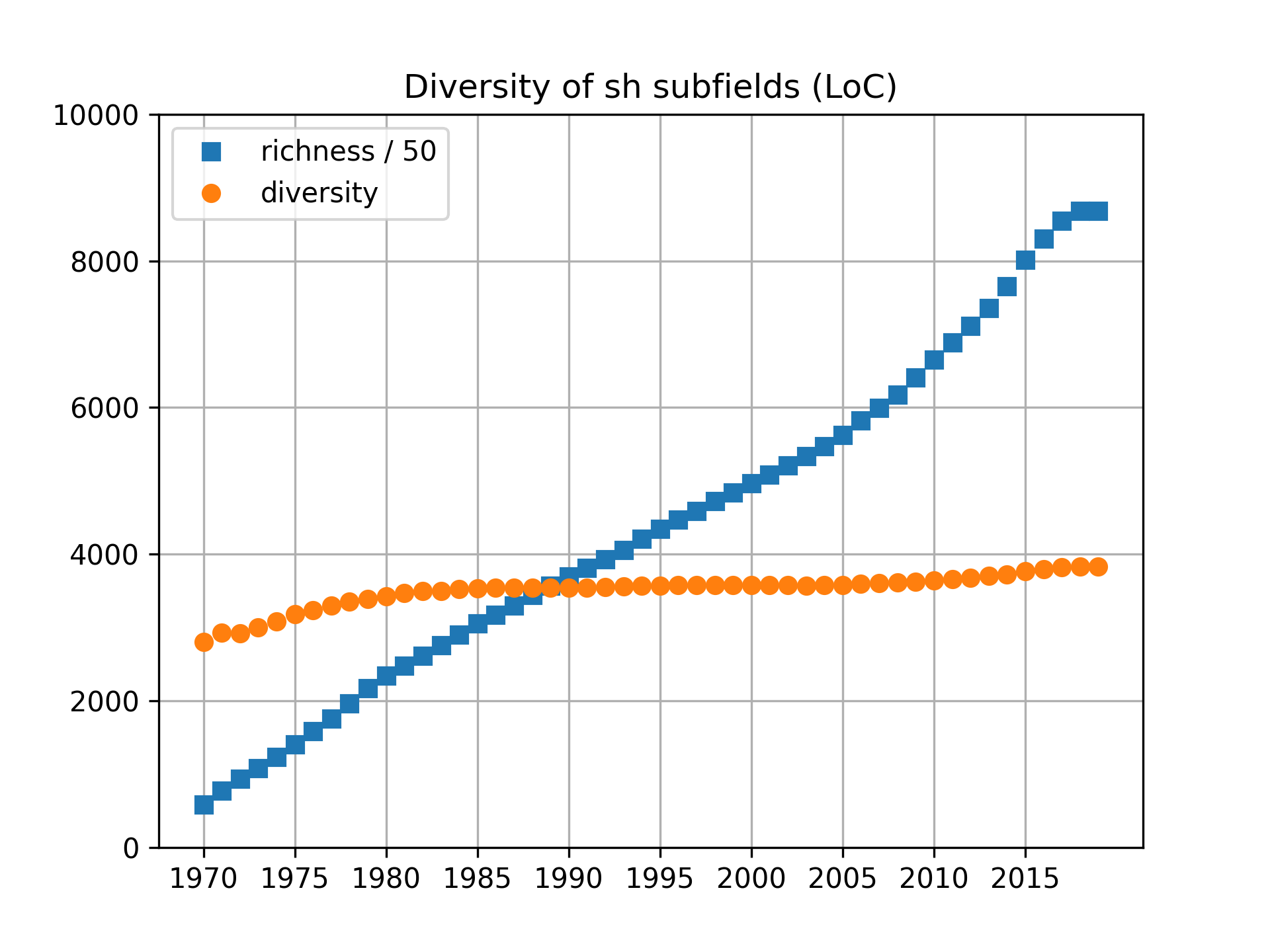}
  \hfill
  
  \includegraphics[width=0.45\textwidth]{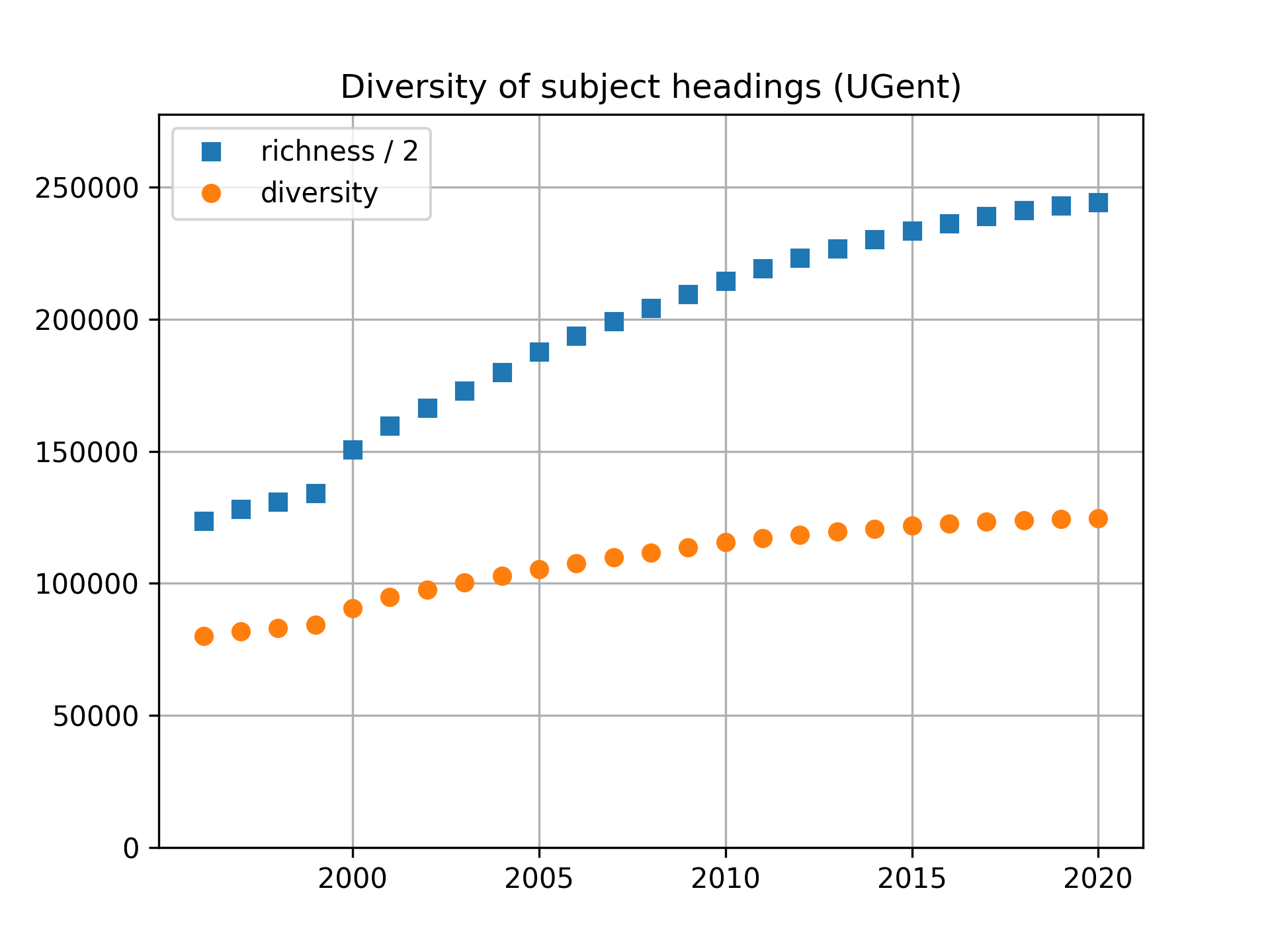}
  \includegraphics[width=0.45\textwidth]{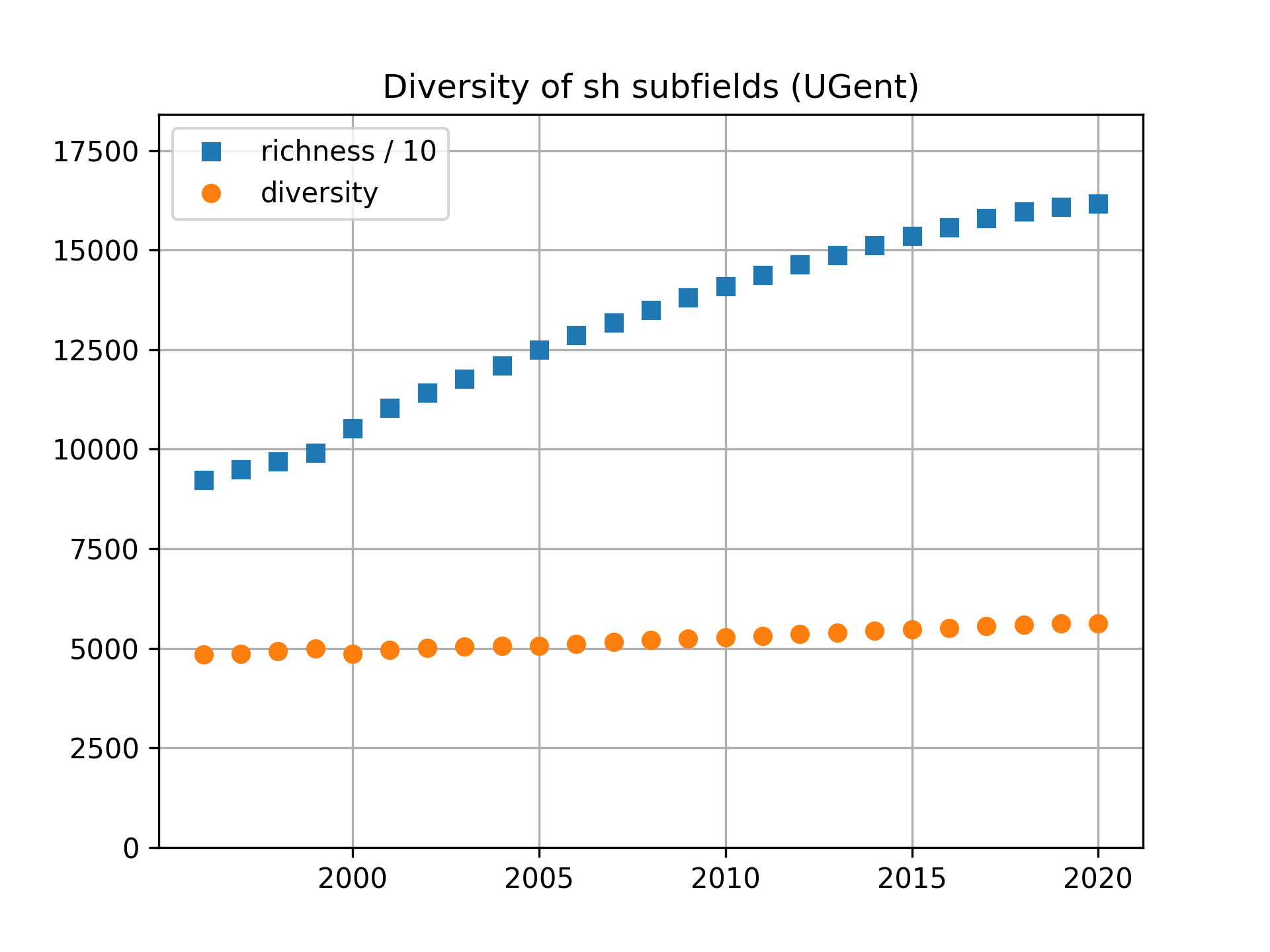}
  \hfill
  \includegraphics[width=0.45\textwidth]{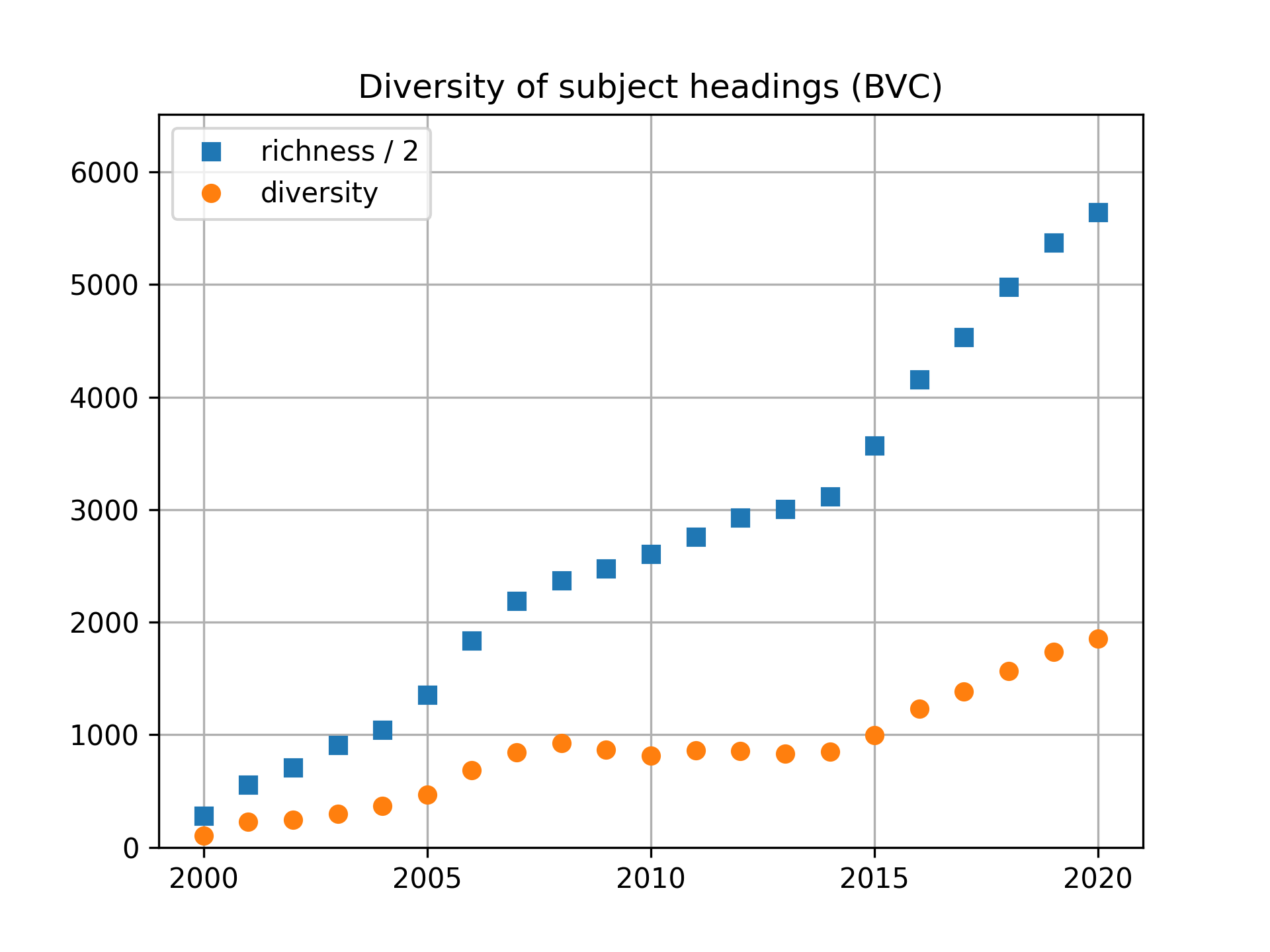}
  \includegraphics[width=0.45\textwidth]{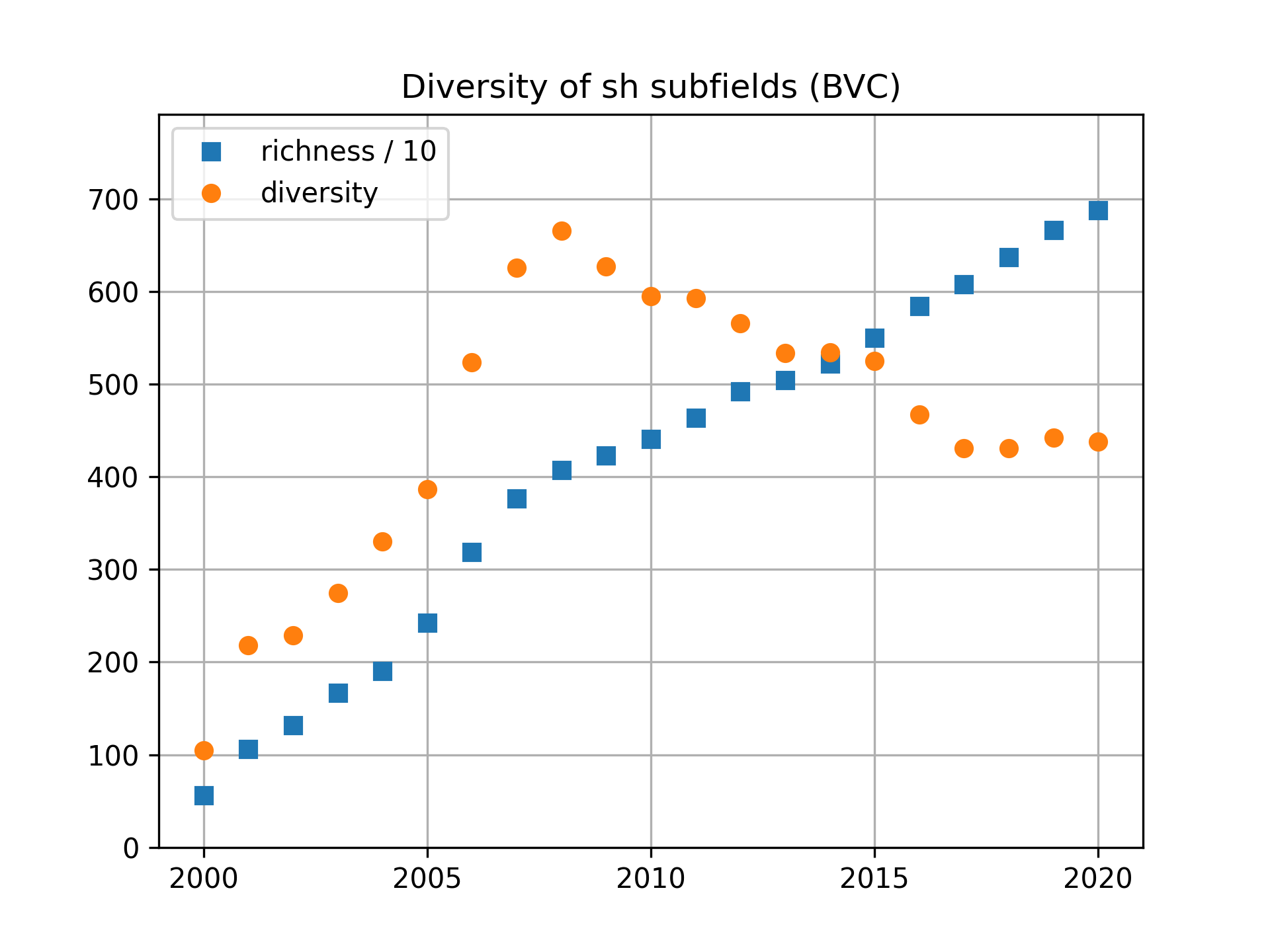}
  \caption{Cumulative richness and Shannon diversity index of the subjects in the catalog. Left: complete subject headings. Right: subject heading subdivisions. Note the specific scales used for richness.}
  \label{fig:subject_headings}
\end{figure*}

We also investigated whether the coverage of topics in a
digital library remains stable, serving a specialized audience, or whether it tends to cover a wider spectrum.  Figure~\ref{fig:subject_headings} shows the trends when the complete descriptor of the subject heading field is analyzed and when its content is split into topical, chronological, geographical, or other subdivisions (so that, for example, the descriptor Commerce--History becomes two subjects, Commerce and History).

In the samples analyzed, the variety of subjects typically shows a constant growth with time, both in terms of richness and diversity. However, this is not the case for the BVC library when the subjects are decomposed into subdivisions. This is due, on the one hand, to a more intensive usage of chronological subdivisions.  On the
other hand, an inspection of the records reveals that the library has, after an initial period, progressively increased the fraction of content within the fields of history and literature (and, remarkably, theater) in Spanish---which now account for nearly one third of its content. The BVC has thus recently developed into a more specialized library.

\subsection{Linked open data}
\label{sec:LOD}
Over the last decade, cultural heritage institutions have moved towards
adopting the \emph{semantic web}~\cite{BernersLee2001} and
\emph{linked open data} concepts by using the W3C \emph{Resource Description Framework} to express semantic relationships~\cite{RDF} and the SPARQL~\cite{SPARQL} language to query them. RDF describes \emph{resources} (the content of a library) by categorizing them in \emph{classes} (such as person, work or name) and uses
\emph{properties} (such as author) to express relationships between resources. Both resources and properties are identified by URIs (Uniform Resource Identifiers): for example, a triple $(X, P, Y)$ can
link the identifier of a person $X$ to the identifier of a name $Y$ connected by the property $P$, where the meaning of URI $P$ is \emph{has name}. Analogously, a triple of the form $(X, \mathtt{rdf:type}, Z)$ declares $X$ to belong to class $Z$.

Libraries have progressively adapted their catalogs~\cite{Smith-Yoshimura2020} to facilitate the publication of Linked Open Data (LOD) repositories. As shown in Table~\ref{tab:DL_LOD}, they have used a variety of vocabularies for the definition of RDF classes and properties, however. The repositories have also been made available in various forms, which include public SPARQL endpoints, OAI-PMH interfaces and even open-access dump
files.\footnote{\url{http://www.openarchives.org/pmh}}

\begin{table*}[h]
  \small\centering
  \caption{Linked Open Data repositories published by libraries.}
  \label{tab:DL_LOD}
  \begin{tabular}{p{0.4\linewidth}p{0.20\linewidth}l}
    Institution & Vocabularies & URL
    \\\toprule
 
    Austrian National Library
                & \textsc{edm} \textsc{bibframe} \textsc{rda}
                               & \url{labs.onb.ac.at/en/dataset/lod}
    \\
    Biblioteca Nacional de España
                & \textsc{frbr}
                               & \url{datos.bne.es}
    \\
    Biblioteca Virtual M. de Cervantes
                & \textsc{rda} &
                                 \url{data.cervantesvirtual.com}
    \\
    Bibliothèque nat. de France
                & \textsc{frbr} & \url{data.bnf.fr}
    \\
    Bibliothèque nat. du Luxembourg
                & \textsc{xml} &
                                 \url{data.bnl.lu}
    \\
    British National Bibliography
                & \textsc{bibo} & \url{bnb.data.bl.uk}
    \\
    Europeana & \textsc{edm} & \url{pro.europeana.eu/page/sparql}
    \\
    Deutsche Nationalbibliothek
                &  \textsc{bibframe} &
                                       \url{www.dnb.de/EN/lds}
    \\
    Library of Congress
                &  \textsc{bibframe} &
                                       \url{id.loc.gov}
    \\
    National Library of Finland 
                & Schema.org  \textsc{bibframe}
                               & \url{data.nationallibrary.fi}
    \\
    Koninklijke Bibliotheek
                & Schema.org  \textsc{lrm}
                               & \url{data.bibliotheken.nl}\\
    \bottomrule
\end{tabular}
\end{table*}

In order to test the application of diversity indices to LOD, data shown in Table~\ref{tab:DL_LOD} were retrieved from these repositories which distribute them with open licenses and via a public SPARQL endpoint. We note that these endpoints may not always reflect the current
situation of the libraries.\footnote{For example, as of March 2022, the Europeana SPARQL endpoint has not been updated since July 2017.} The harvesting was performed with simple scripts,\footnote{Some repositories implement a timeout limit for the downloads. In such cases, partitioned queries were needed to retrieve all the information.} such as those presented in Appendix~\ref{sec:sparql}.

\begin{table*}[h]
  \centering
  \begin{tabular}{l|rrr|rrr}
    Resource type & \multicolumn{3}{|c|}{class} & \multicolumn{3}{c}{property}
    \\
    host & $D$ & $R$ & $D/R$ & $D$ & $R$ & $D/R$ \\\hline
    AT & 2.1 & 5 & 0.42 & 10.7 & 22 & 0.48 \\
    BNB & 13.2 & 33 & 0.40 & 26.6 & 88 & 0.30 \\
    BNE & 3.8 & 16 & 0.24 & 50.9 & 189 & 0.27 \\
    BNF & 6.9 & 26 & 0.27 & 55.5 & 791 & 0.07 \\
    BVC & 6.6 & 27 & 0.24 & 32.0 & 165 & 0.19 \\
    EU & 5.1 & 11 & 0.46 & 37.1 & 115 & 0.32 \\
    FI & 7.0 & 12 & 0.59 & 17.3 & 35 & 0.49 \\
    KB & 3.9 & 12 & 0.32 & 14.6 & 23 & 0.64 \\
    \hline
  \end{tabular}
  \caption{Diversity $D$, richness $R$ and diversity-richness rate $D/R$ of the resources contained in linked open data.}
  \label{tab:LOC_resources}
\end{table*}

The diversity $D$ and richness $R$ of the resources was computed, as well as the diversity to richness ratio, which provides an indication of how effective the usage of the available tags is. As shown in Table~\ref{tab:LOC_resources}, some libraries, such as the
Austrian National Library (AT), the National Library of Finland (FI) and the Koninklijke Bibliotheek (KB) employ vocabularies with a small number of classes and properties. In contrast, the National Library of France (BNF) and the National Library of Spain (BNE) describe their resources in terms of the richer FRBR and RDA vocabularies. The BNF also employs a proprietary vocabulary to describe the roles of creators which contains over 500 categories.  Since they are not uniformly used, this leads to a lower $D/R$ ratio. The British National Bibliography (BNB) is an intermediate case, as it essentially employs the BIBO vocabulary which contains 33 classes and 88 properties.

\begin{figure*}
  \centering
  \includegraphics[width=0.8\textwidth]{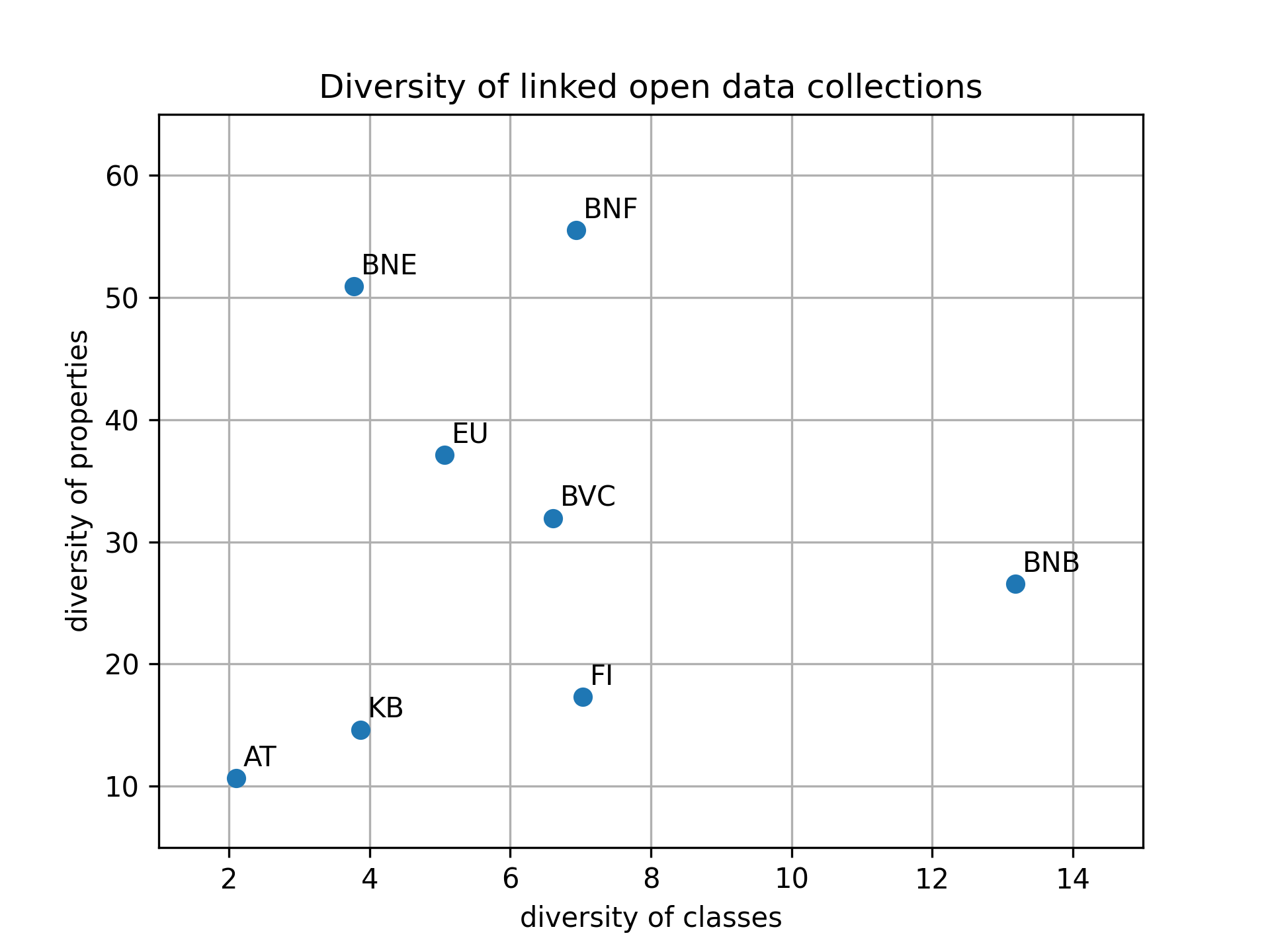}
  \caption{Shannon diversity of classes and properties in \emph{linked open data} published by libraries.}
  \label{fig:LOD}
\end{figure*}

Although there is a moderate positive correlation between the diversity of classes and the diversity of properties employed in each collection (see Figure~\ref{fig:LOD}), some libraries show a finer granularity of classes while other employ a higher variety of properties.

\section{Conclusions}
\label{sec:conclusions}
Diversity indices provide a complementary view of the variety of the
groups in a collection of data. In contrast to richness, diversity is
more robust than the sample size as it gives less weight to classes with
a smaller number of occurrences.

When lexical content is analyzed, the diversity of words approaches an
asymptotic value which depends on the author and genre of the
works. This value can be obtained by extrapolating the observed values
with a simple model involving only three free parameters. The
extrapolation proves stable with respect to the size of the sample.

As regards metadata, diversity indices can be used to visualize the
trends, for example, in creator or subject coverage. The rate between
diversity and richness also proves useful to compare the effective
usage of the available descriptors (classes and properties) to
describe resources in the semantic data (linked open data collections)
published by digital libraries.

The Python scripts employed for the analysis included in this paper
have been published as open-access software in~\cite{Carrasco2022}.

\bmhead{Acknowledgments}
We thank Frank Vandepitte and Patrick Hochstenbach from the Ghent University Library for their kind assistance in understanding the library catalographic records.

\begin{appendices}

\section{SPARQL queries}\label{secA1}
\label{sec:sparql}

\begin{lstlisting} [
  language=SQL,
  label=query,
  frame=tb,
  % 
  caption=Query used to retrieve all classes
  and the number of resources per class in a LOD repository.
  %
  ]
SELECT ?class (COUNT(?s) AS ?count)
WHERE {
    ?s a ?class 
}
GROUP BY ?class
\end{lstlisting}

\begin{lstlisting} [
  language=SQL,
  label=query-sameas,
  frame=tb,
  % 
  caption=Query retrieving external repositories linked
  from a specific LOD repository and the number of links to each one.
  %
  ]
SELECT ?hostname (COUNT(?s) AS ?count)
WHERE{
   ?s owl:sameAs ?same . 
    bind(
       strbefore(strafter(
        str(?same),"//"),"/") 
        AS ?hostname)
}
GROUP BY ?hostname
\end{lstlisting}

\end{appendices}


\bibliography{diversity-bibliography}


\end{document}